\documentclass[aip,jcp,
 amsmath,amssymb,
 reprint,%
floatfix]{revtex4-2}

\usepackage{graphicx}
\usepackage{dcolumn}
\usepackage{bm}
\usepackage{siunitx}
\usepackage[utf8]{inputenc}
\usepackage[T1]{fontenc}
\usepackage{mathptmx}
\usepackage{etoolbox}
\usepackage{float}
\usepackage{booktabs}
\usepackage{multirow}
\usepackage{color}
\usepackage{soul}
\usepackage{ulem}
\usepackage{todonotes}
\usepackage{silence}
\usepackage{array}
\newcolumntype{P}[1]{>{\centering\arraybackslash}p{#1}}

\DeclareSIUnit\angstrom{\text {Å}}

\makeatletter
\def\@email#1#2{%
 \endgroup
 \patchcmd{\titleblock@produce}
  {\frontmatter@RRAPformat}
  {\frontmatter@RRAPformat{\produce@RRAP{*#1\href{mailto:#2}{#2}}}\frontmatter@RRAPformat}
  {}{}
}%
\makeatother
\begin{document}

\preprint{AIP/123-QED}

\title{Probing Water-Electrified Electrode interfaces: Insights from Au and Pd}

\author{Graciele M. Arvelos}
\affiliation{Instituto de F\'isica Te\'orica, Universidade Estadual Paulista (UNESP), S\~ao Paulo, SP 01140-070, Brazil}

\author{Marivi Fernández-Serra}
\affiliation{Physics and Astronomy Department, Stony Brook University. Stony Brook, New York 11794-3800, USA}
\affiliation{Institute for Advanced Computational Science, Stony Brook, New York 11794-3800, USA}

\author{Alexandre R. Rocha}
\affiliation{Max Planck Institute for the Structure and Dynamics of Matter, 22761 Hamburg, Germany}
\affiliation{Instituto de F\'isica Te\'orica, Universidade Estadual Paulista (UNESP), S\~ao Paulo, SP 01140-070, Brazil}

\author{Luana S. Pedroza}
\affiliation{Instituto de F\'isica, Universidade de S\~ao Paulo, SP 05508-090, Brazil}
\email{luana@if.usp.br}

\date{\today}

\begin{abstract}

The water/electrode interface under an applied bias potential is a challenging out-of-equilibrium phenomenon, which is difficult to accurately model at the atomic scale. In this study, we employ a combined approach of Density Functional Theory (DFT) and non-equilibrium Green’s function (NEGF) methods to analyze the influence
of an external bias on the properties of water adsorbed on Au(111) and Pd(111) metallic electrodes. 
Our results demonstrate that while both Au and Pd-electrodes induce qualitatively similar structural responses in adsorbed water molecules, the quantitative differences are substantial, driven by the distinct nature of water-metal bonding. Our findings underscore the necessity of quantum-mechanical modeling for accurately describing electrochemical interfaces.

\end{abstract}

\maketitle

\section{\label{sec:introduction} Introduction}

An atomistic description of the water-metal interface is important for a better comprehension of a myriad of processes such as heterogeneous catalysis,\cite{catalysis,catalysis2,catalysis3,catalysis4,review_edl} corrosion resistance,\cite{corrosion} and catalytic processes in solar cells.\cite{hydrogen_electrode,electro_curcinotta,electro_review}
In this context, many properties that characterize the reactivity and electrochemical behavior of the interface are ultimately defined by the electronic response to external factors and perturbations, such as an applied external potential.\cite{michaelides_review} These properties are related to atomic arrangements determined by the orientations of water molecules, the formation of different ions at the interface, and the reactions that might occur at different potentials.\cite{bias-pd}

An electrochemical cell can be modeled  by two independent charge reservoirs acting as electrodes separated by an electrolytic solution. Applying an electrostatic potential difference across such a device will lead to a redistribution of charge both at the electrode interface as well as in the ions composing the electrolyte. At the microscopic level, this process results in the formation of the electric double layer (EDL), responsible for the formation and breaking of chemical bonds, as well as possible charge transfer processes and adsorption occurring at the electrode interface.\cite{electro_curcinotta,edl_review} However, even for concentrated electrolytes, water molecules dominate this interface and understanding the water-metal interaction is a first step towards a better comprehension of the EDL.\cite{bias_agua1} The microscopic structure of water at the metal interface can be inferred from vibrational spectroscopy techniques combined with atomistic modelling, due to the correlation between frequencies and hydrogen bond strength.\cite{sfg_review,vibrational_review,sfg_review,rx1,rx2,sfg1,sfg2,sfg3,sfg_kramer,raman1,raman2,experimental_pd,experimental_gold}

Computer simulations have been crucial in analyzing and describing these structures at the atomic level and assisting in the interpretation of experimental results.\cite{experimental_gold} In particular, density functional theory (DFT) has elucidated experimental questions, such as the stability of water structures adsorbed on transition metals.\cite{Poissier2011,LSP-PdH2O,michaedelis,meng_water_adsorption} However, modeling an electrified electrode/electrolyte interface with atomistic accuracy and at a reasonable computational cost is still a major challenge.\cite{electro_review,review_electrode,electrochemical,double_layer,modeling_electrochemical,negf,swiftModelingElectricalDouble2021} In the realm of DFT, \citeauthor{bias-pd}\cite{bias-pd} implemented the electrification of a metal surface by modifying the total number of electrons on the electrode. Using this approach, the authors studied the response of the polarization of water/Pd(111) and water/Cu(111) interfaces to an external potential.\cite{charge2,bias-pd} Other implementations in this direction have emerged in the literature over the years, providing details about the water-metal interface, polarization effects, insights into processes occurring in the EDL, and information about activation energies.\cite{hydrogen2,hydrogen1,hydrogen3,hydrogen4,hydrogen5} In all these studies, the variable that is theoretically controlled is the added (or subtracted) charge to/from the electrode, and the resulting applied potential is then inferred from the computed average electric field in the device. This does not replicate experiments, where the applied potential is the control variable, and the electrode charges depend on the response of the electrolyte to such applied bias. There were also developments in the direction of controlling the potential, using a potentiostat. However, in this approach the instantaneous potential and dynamics are not described, and what is obtained is the average potential at the electrode. \cite{PhysRevLett.109.266101,PhysRevLett.126.136803}

Recently we have proposed\cite{negf_gold} a methodology that correctly models the experimental scenario. 
It is based on the non-equilibrium Green’s function (NEGF) formalism combined with DFT.\cite{Datta_1995,smeagol1,transiesta1,transiesta2}
It is specifically designed to system under an external bias potential and can be readily applied to a solid-liquid interface. It has the advantage of controlling the bias considering the electrodes as semi-infinite reservoirs,\cite{sanvito,negf,negf_gold} as well as obtaining the forces in an out-of-equilibrium situation with ab-initio precision.\cite{force1,force2,negf_cp2k} In this work, we apply this methodology to describe the effect of electrically biased surfaces on water structures (monomer and a monolayer) adsorbed on two distinct metal surfaces -- Au(111) and Pd(111). This allows us to quantitatively evaluate the differences between these two electrodes.

\section{\label{sec:method}Computational Details}

\begin{figure}[H]
	\includegraphics[width=\columnwidth] {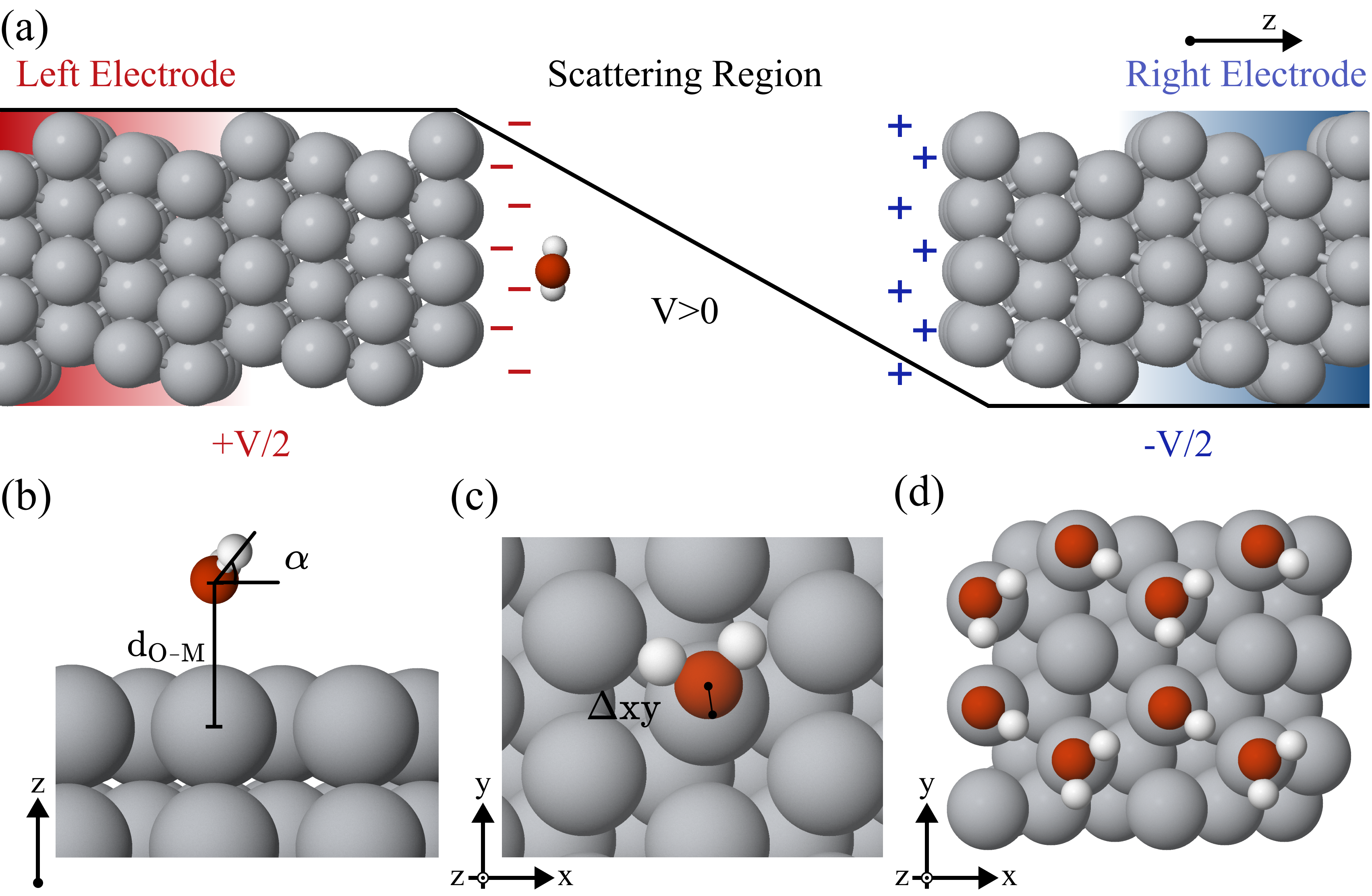}
\caption{(a) Illustration of the electrochemical setup in the framework of the NEGF formalism used to apply an external bias potential into the water-metal system. The LE and RE regions correspond to the left and right electrodes, respectively, and the SR is known as the scattering region; (b) side and (c) top views of a water monomer on the metallic surface and structural properties measured after optimization: the vertical distance between oxygen and metal atom ($\rm{d_{O-M}}$), tilt angle between metal surface and the plane formed by the molecule ($\rm{\alpha}$), and the horizontal displacement from the top position on the metal surface ($\rm{\Delta xy}$), respectively; (d) top view of the H-down water layer.}
	\label{fig:system}
\end{figure}

For studies of the isolated water monomer, we have analyzed its structural and vibrational dependence upon bias changes in two different surfaces, Au(111) and Pd(111).
We made this choice because they expand the range of water-adsorption mechanisms in noble metals. While for Au the adsorption is
dominated by electrostatics, in Pd a significant amount of charge transfer at the interface occurs.\cite{LSP-PdH2O,Poissier2011} 
In addition we also analyze the behavior of a monolayer of 2D ice water adsorbed on the Au electrode. 
The systems considered in our simulations were composed of two semi-infinite metal slabs acting as electrodes, and the water (monomer and monolayer) in contact with one of them, as illustrated in Fig.\ref{fig:system}(a). This is similar to the arrangements used in standard electronic transport calculations. For the simulations, the system is split into three regions\cite{caroli}, namely two semi-infinite electrodes in thermal equilibrium - albeit with a possibly different chemical potential if a bias is applied - and a so-called scattering region (SR), which consists of a water molecule/monolayer and a number of layers of metal on either side. The number of layers used ensures that the charge density at both edges of the simulation box is the same as the one deep inside the bulk electrodes.

For the gold electrodes (both for a water monomer and for a monolayer), the optimized bulk lattice constant is 4.24 \si{\angstrom} and the system was constructed with 3 layers, each one containing $3\times4$ Au atoms on the surface plane forming the leads on either side. These are then attached to four layers on the left and three layers on the right forming the full simulation cell as illustrated in Fig. \ref{fig:system}(a). The left and right metallic surfaces are separated by 25 \si{\angstrom}, and the water molecule or the 2D monolayer adsorbed on the left electrode. We have chosen a H-down layer as seen in Fig. \ref{fig:system}(d), as this was shown to be a stable configuration.\cite{meng_water_adsorption} The layer consisted of 8 water molecules in a hexagonal arrangement, similar to ice Ih, in a unit cell of size $\rm{\sqrt{3}\times \sqrt{3}R30}$ relative to the lattice parameter of Au, as this makes them almost commensurate (Fig. \ref{fig:system}(d)). In the case of the monomer adsorbed on Pd electrodes, the Pd bulk lattice constant is $3.97$ \si{\angstrom}, and again we used a layer of Pd formed by 3$\times$4 atoms. The electrodes now consist of 6 layers to guarantee coupling between neighboring cells only. These are again connected to four and three layers to the left and right, respectively, and the vacuum region is kept at 25 \si{\angstrom}.

Initially we performed standard equilibrium density functional theory calculations with periodic boundary conditions (PBC) using the Siesta\cite{siesta} code with PBE\cite{pbe} as our choice of exchange and correlation potential (XC). The valence electrons were described by optimized numerical atomic orbitals with double-$\zeta$ polarization,\cite{LSP-PdH2O,negf_gold} and the core electrons were described by norm-conserving pseudopotentials in the Troullier-Martins form\cite{troullier}. Subsequently, a finite voltage was applied to the electrodes, shifting the left (right) chemical potentials by $\rm{V/2}$ ($\rm{-V/2}$) and inducing a negative (positive) charge on the surface for positive (negative) bias. Thus, the problem becomes a non-equilibrium one and the calculations were performed self-consistently using the NEGF formalism coupled with DFT with the same parameters used for the ground state calculations. For each applied bias the system is relaxed by self consistently calculating the non-equilibrium forces.\cite{force1,force2,negf_gold}
All non-equilibrium calculations were performed using the Transiesta code.\cite{transiesta1,transiesta2}

The minimum energy configuration of the water monomer adsorbed on each metal was obtained with the criteria of $0.005\;\si{\eV}/\si{\angstrom}$ for the maximum force on each atom and with the electrode atoms fixed at the bulk geometry (the metal atoms closer to the surface are allowed to move for PBC optimization). The final configuration is then used as a starting point for the next bias voltage. Using the optimized geometry, the vibrational modes for the water molecule were obtained by diagonalizing a numerically computed force constant matrix where the water atoms are displaced by 0.01 Bohr from their equilibrium positions. Parameter convergence in structural optimizations incorporating NEGF methods is critical to ensure the validity of the observed simulation results. In particular, one needs to ensure that the number of metal-electrode layers is sufficient. At zero applied bias, structural properties of adsorbents
computed using a n-layers metal slab with PBC should not differ from those obtained in a NEGF simulation with semi-infinite separated electrodes. We have ensured that our results are converged using these criteria.

\section{\label{sec:result} Results and Discussion}

\subsection{Water monomer}

\begin{figure*}[hbt]
	\includegraphics[scale=.062]{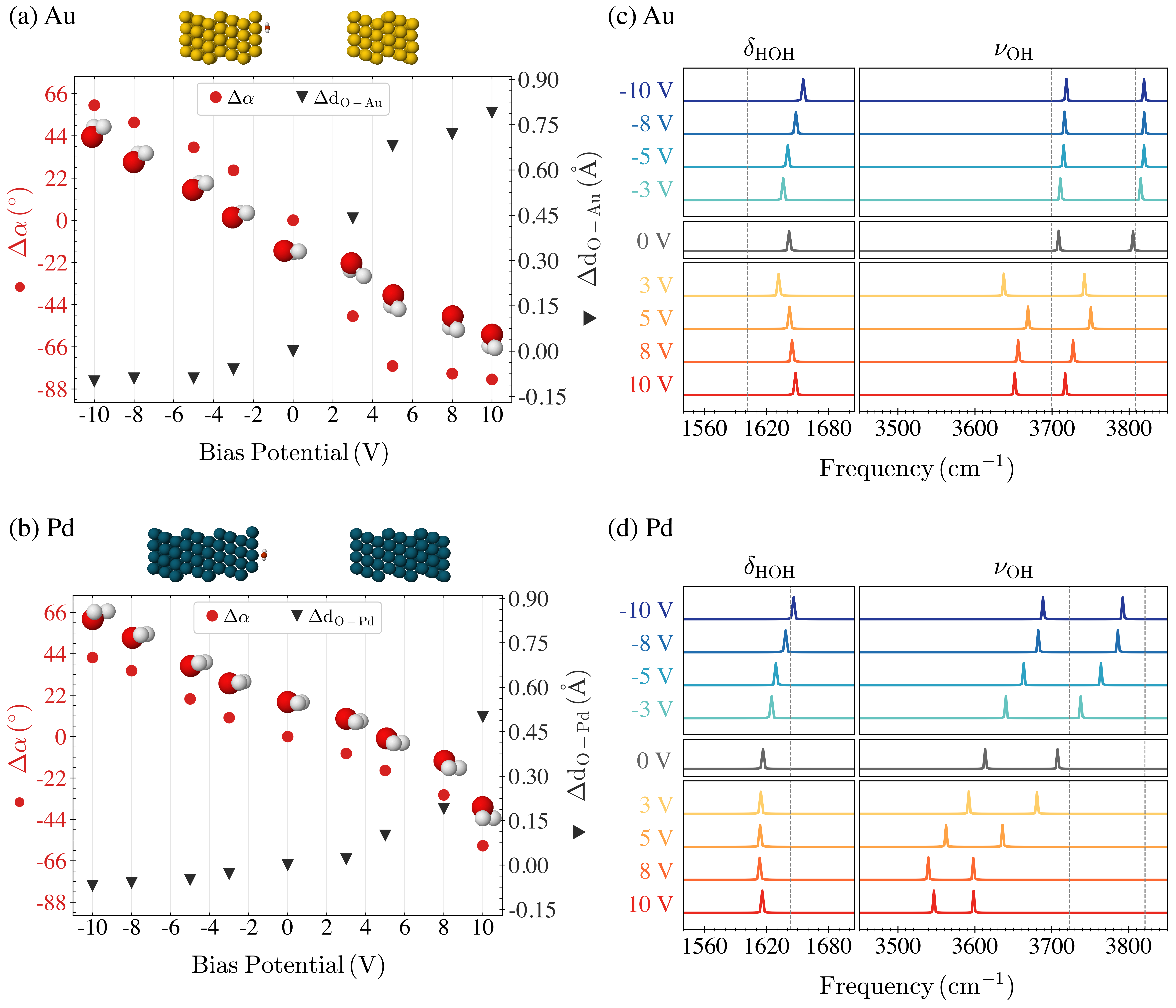}
\caption{Relative tilt angle and water-metal distance of a water monomer adsorbed on (a) Au(111) and (b) Pd(111) surfaces as a function of applied bias, compared to the zero-bias configuration. The bias-dependent equilibrium water configuration is also shown. (c) and (d) show the normal mode frequencies of a water molecule adsorbed on Au(111) and Pd(111) surfaces under applied bias, respectively. The dotted line represents the frequencies of an isolated water monomer, calculated with the inclusion of metal ghost atoms.}
    \label{fig:monomer}
\end{figure*}

In the absence of applied potential, for both Pd and Au surfaces, the optimal arrangement for the water monomer adsorbed on it is a flat configuration, \cite{dft_water_metals,role_vdw,Poissier2011,negf_gold,meng_water_adsorption,michaedelis,charge_density_monomer} {\it i.e.} the water dipole is approximately parallel to the surface as measured by $\alpha$, the angle between the dipole moment and its component on the $\rm{xy}$ plane (Fig. \ref{fig:system}(b) and (c)).
For Au, the bonding between molecule and surface is weaker, and the distance to the oxygen ($\rm{d_{O-M}}$) atom is $2.80\;\si{\angstrom}$, compared to Pd which is $2.47\;\si{\angstrom}$, obtained using standard periodic DFT calculations. These values are close to the values ($2.83-3.14)\;\si{\angstrom}$ for Au and ($2.37-2.47)\;\si{\angstrom}$ for Pd reported in the literature using the same XC functional.\cite{dft_water_metals,role_vdw,Poissier2011,negf_gold}
We also analyzed the effects of including van der Waals corrections with the VDW-BH\cite{vdw_bh} exchange and correlation functional for Pd electrodes, and we obtained $2.39\;\si{\angstrom}$ for $\rm{d_{O-M}}$ and $2^{\circ}$ for the tilt angle, revealing results closer to the PBE XC.\cite{role_vdw,LSP-PdH2O,Poissier2011} Therefore, for the calculations applying an external potential bias, we will use only PBE.
We also minimized the configuration using the NEGF-DFT methodology considering $\rm{V}=0\;\si{\V}$, as can be seen from Table \ref{tab:monomer_properties} for the metal-oxygen distance and the tilt angle $\alpha$. We note that these structural properties did not change compared to the PBC values. This indicates that the amount of metal included is sufficient for screening.

\begin{table*}[t!]
\caption{Structural properties and normal mode frequencies values of a monomer adsorbed on Pd(111) and Au(111) surface according to the applied biases. We present the water-metal distance ($\rm{d_{O-M}}$), the water molecule tilt angle $\alpha$ with respect to the metal surface plane, the horizontal displacement in comparison with the \textit{atop} position along the $\rm{xy}$ plane ($\Delta \rm{xy}$), the frequencies of the bending normal mode ($\rm{\delta_{HOH}}$), and symmetric ($\rm{\nu_S}$) and antisymmetric ($\rm{\nu_{AS}}$) modes.\label{tab:monomer_properties}}
\begin{ruledtabular}
\begin{tabular}{ccccccccccccc}
V (\si{\V}) & \multicolumn{2}{c}{$\rm{d_{O-M} ~ (\si{\angstrom})}$} & \multicolumn{2}{c}{$\alpha ~(\si{\degree})$} & \multicolumn{2}{c}{$\rm{\Delta xy
~ (\si{\angstrom})}$} & \multicolumn{2}{c}{$\delta_{\rm{HOH}} ~(\si{\cm}^{-1})$} & \multicolumn{2}{c}{$\nu_{\rm{S}} ~(\si{\cm}^{-1})$} & \multicolumn{2}{c}{$\nu_{\rm{AS}} ~(\si{\cm}^{-1})$}  \\ 
\midrule
 & Au  & Pd                                      & Au   & Pd                              & Au      & Pd                                & Au      & Pd             & Au & Pd                & Au   & Pd              
 \\ \midrule
-10                                   &   2.71                              & 2.39                             & 56                     & 38                     &       0.24               & 0.07                 &  1655               & 1645          & 3718      & 3689    & 3820     & 3793                 \\
-8                                    &      2.72                              & 2.40                             & 47                     & 31                     &        0.17              & 0.07                 &    1648           & 1637          &   3717  & 3683    &  3821    & 3787                 \\
-5                                    &    2.72                              & 2.41                             & 34                    & 16                     &        0.10               & 0.24                 &     1640           & 1630          &   3715     & 3663    &   3820    & 3765                 \\
-3                                    &  2.75                              & 2.43                             & 22                    & 6                      & 0.02                          & 0.38                 &    1636           & 1624          &    3712    & 3640    &    3816  & 3738                 \\
0                                     &  2.81                              & 2.46                             & -4                   & -4                     & 0.35                          & 0.49                 &    1642          & 1616          &    3708    & 3614    &   3806   & 3708                 \\
3                                     &  3.25                             & 2.51                             & -54                    & -13                    & 1.12                         & 0.55                 &        1633        & 1614          &  3638   & 3592    & 3743     & 3683                 \\
5                                     &  3.49                              & 2.56                             & -80                    & -22                    & 1.38                          & 0.67                 &      1642          & 1614          &   3669    & 3552    &   3751    & 3637                 \\
8                                     &  3.53                              & 2.65                             & -84                   & -35                    & 1.42                          & 0.77                 &      1646         & 1614          &  3657   & 3540    &   3728  & 3599                 \\
10                                     & 3.60                             & 2.96                             & -87                    & -62                    & 1.55                          & 1.02                 &    1648            & 1616          &   3652    & 3547    & 3718     & 3599                \\ \midrule
\multicolumn{7}{c}{Isolated monomer (Metallic Ghost Atoms)}  &    1602              &      1643           &    3699            &       3723        &        3808       &       3821        \\ 
\end{tabular}
\end{ruledtabular}
\end{table*}

Starting from the flat configuration we applied a positive/negative external potential bias, up to $10\;\si{\V}$ (as mentioned before, the effective potential seen on any metal surface corresponds to $\rm{V/2}$ in the NEGF formalism). In Figs. \ref{fig:monomer} (a) and (b) we present the final relaxed water structure configuration as a function of the applied bias potentials as well as the distance between the metal and the oxygen, and the tilt angle for each electrode.
This information is also presented in a more quantitative form in Table \ref{tab:monomer_properties}.

In general, for both metals we observe a tendency for the oxygen (hydrogen) atom to approach (move away) the surface for $\rm{V} < 0$ and to move away (approach) from it for $\rm{V}>0$. 
This occurs due to the polarization of negative (positive) surface charge on the metal that arises when applying a positive (negative) bias to the electrodes.\cite{negf_gold} 
In accordance with this behavior, previous studies have observed a deviation of the flat configuration to an up (down) configuration for positive (negative) metal surface polarization for Au electrodes\cite{negf_gold} with NEGF formalism, as well as for a homogeneous electric field applied to a water molecule on a Pd slab.\cite{bias_pd_monomer}.

It is important to note that the effect of the potential is not symmetric with respect to polarity. Under a positive potential bias, the tilt angle is larger compared to that under a negative bias, which can be attributed to differences in charge transfer. As can be seen in Fig. \ref{fig:monomer}(a) and Table \ref{tab:monomer_properties} 
the structural properties of the gold electrodes are more significantly impacted due to the weaker bonding between the water molecule and the gold electrode. 
For example, at a relatively low bias ($\sim 5$ V), the water molecule rotates into a downward configuration (with hydrogens pointing towards the metal), and the oxygen-metal distance ($\rm{d_{O-M}}$) shows minimal change as the potential increases. In contrast, on the Pd surface, the water molecule is more strongly bound, with the downward configuration occurring only at higher voltages ($10\, \rm{V}$). Additionally, we observed a shift in the preferred adsorption site with positive potential, as indicated by the horizontal displacement of the water molecule ($\Delta xy$). For negative bias potentials, the water favors the \textit{atop} position, while for positive bias potentials, the \textit{hollow} site becomes preferred.

\begin{figure}[b!]
    \centering
    \includegraphics[width=.82\linewidth]{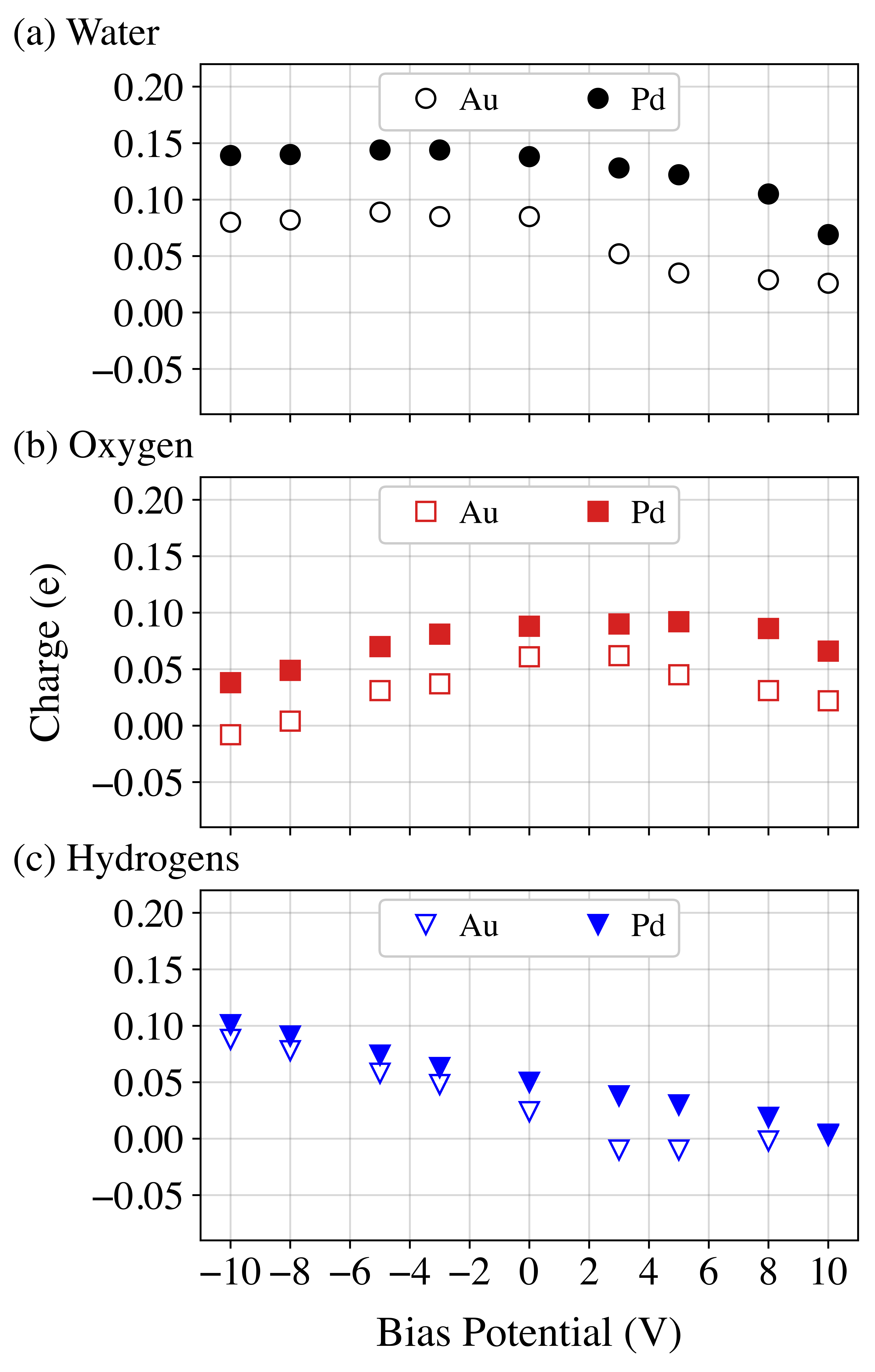}
    \caption{Charge transfer to the water molecule in units of the electron charge for different applied bias potentials measured by Mulliken charge population for Au (empty points) and Pd (filled points) electrodes. (a) The black circles correspond to the total water charge, and the (b) red circles and (c) blue triangles represent the O and H net charge contributions compared to the isolated water monomer.}
    \label{fig:mulliken}
\end{figure}

\begin{figure*}[ht!]
    \centering
	\includegraphics[scale=.15]{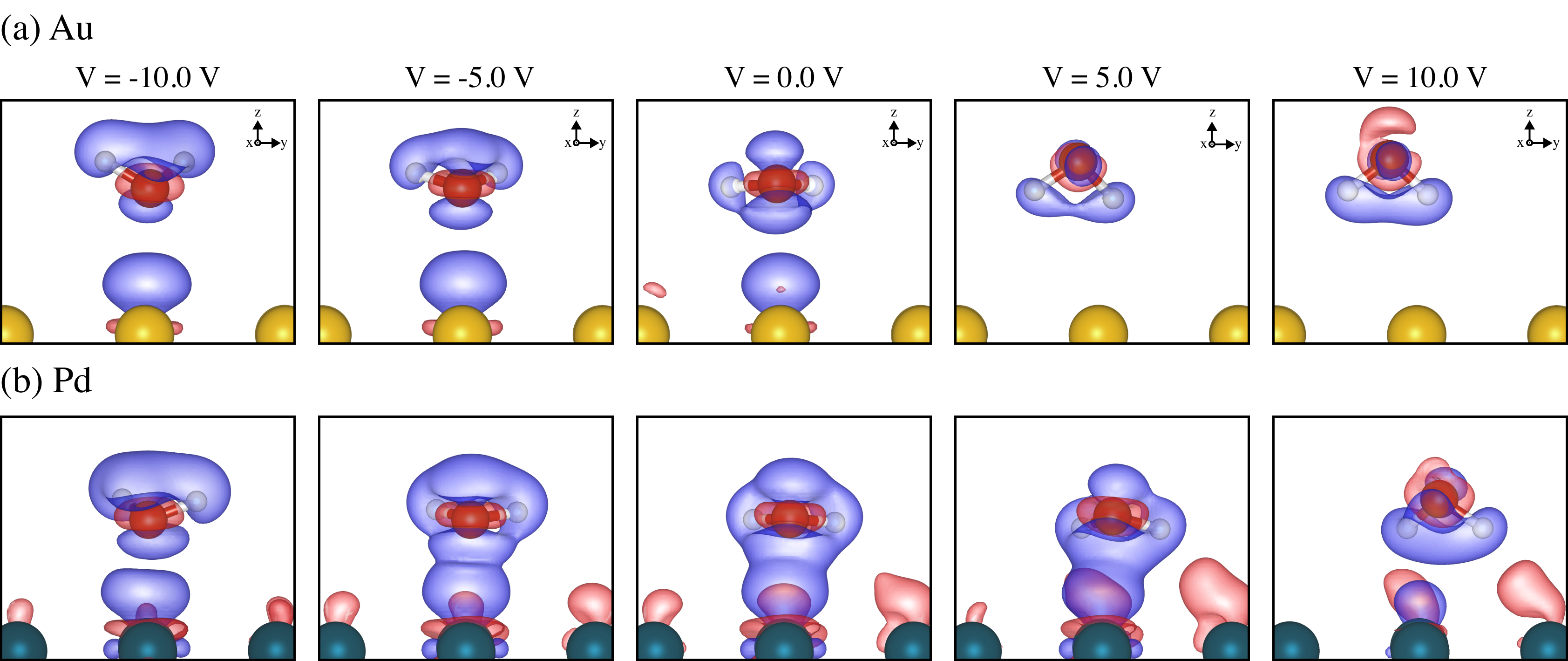}
    \caption{Difference in charge density between the system under bias V and a combined parallel-plate capacitor with the same bias, and an isolated water molecule with an equivalent external electric applied. The isosurface value was $1.5\times10^{-3}\,\si{e}/\si{\angstrom}^3$ for all plots, where red (blue) indicates an excess (depletion) of electrons.}
    \label{fig:chargedifference}
\end{figure*}

To investigate the role of charge transfer in the bias-dependent structural properties, we calculated the overall change in the Mulliken population on the water molecule for each applied bias and compared the atomic charges to those of an isolated water monomer. 
As shown in Figure Figure \ref{fig:mulliken}, Pd exhibits a stronger charge transfer between the metal surface and the water molecule even at zero bias, which accounts for the smaller change in the oxygen-metal distance ($\rm{\Delta d_{O-M}}$) compared to gold. This higher charge transfer in Pd persists across the range of applied biases, both positive and negative.
Moreover, greater charge transfer occurs under negative bias potentials when the water molecule adopts an almost upright configuration and in the \textit{atop} site position, in such way that is favored by the interaction between the oxygen and the metal atoms. 
This behaviour is reflected in the difference in charge density difference between the full system at a given bias and the combination of a parallel plate capacitor with bias V and an isolated molecule in an equivalent electric field (see Figure \ref{fig:chargedifference} and Figures S1 and S2 in the supplemental information file). 
This allows us to analyze the water-metal interactions excluding a classical electrostatic effect. We note that for zero bias we have the characteristic charge density profile associated to Pauli repulsion in molecule-metal systems.\cite{bias_agua1,charge_density_monomer} As the bias negatively increases, 
Pauli repulsion increases as the oxygen atom moves closer to the surface, a competing effect of the the attractive interaction between the positively charged surface and the negatively charged O atom.
It is also interesting to point out that the total charge transfer between surface and molecule is small for negative bias, while there is some degree of charge rearrangement within the water molecule.

The structural changes and charge transfer induced by variations of the external potential can be reflected in the vibrational properties of the system. These have been observed experimentally using surface-enhanced infrared absorption spectroscopy (SEIRAS),\cite{raman1} which showed modifications in the vibrational mode of angular deformation and an increase in the number of hydrogen bonds, for positive voltages. Additionally, sum-frequency generation (SFG) spectroscopy, sensitive only to the water-metal interface, revealed variations in higher-frequency OH stretching due to potential changes, attributing these shifts to free OH bonds oriented towards the gold electrode. They also observed a weak interaction of these molecules with the metal,\cite{sfg1,sfg2,sfg_kramer} as well as a tendency for water molecules to move closer to the electrode under positive potentials and further away under negative potentials, evidenced by an increase in interfacial water density for positive bias.\cite{sfg_kramer}

To gain deeper insight, we first examined the behavior of a single water molecule on the surface. Thus, we have calculated the vibrational modes of the molecule as a function of the applied external potential. In the zero-bias case, the stretching frequencies, $\nu_{OH}$, of the water molecule adsorbed on Au are higher than those on Pd. This difference is due to the weaker binding of the molecule to the Au surface, which keeps the water molecule at a greater distance from the surface compared to Pd. In contrast, the stronger binding to Pd results in lower stretching frequencies, reflecting the closer proximity and stronger interaction between the water molecule and the Pd surface.

In the case of Pd, applying a negative external potential resulted in a significant increase in the antisymmetric stretching frequencies, $\nu_{AS}$, by up to 85 cm$^{-1}$ at $\rm{V} = -10.0,\rm{V}$. In contrast, when the molecule is adsorbed at the gold surface, we observed that $\nu_{AS}$ increased only slightly and then remained essentially constant up to 15 cm$^{-1}$ for negative potentials. This increase in frequencies for both surfaces is due to a reduced interaction between the O-H bonds and the metal, caused by the rotation of the water molecules away from the surface. For positive potentials ($\rm{V} > 0$), the stretching frequencies tend to decrease as the O-H bonds move closer to the metal surface, with a maximum reduction of 109 cm$^{-1}$ for Pd and 88 cm$^{-1}$ for Au. These results are in agreement with those obtained by \citeauthor{bias_pd_monomer}\cite{bias_pd_monomer} who applied an external homogeneous electric field perpendicular to a Pd slab. While a one-to-one correspondence between field and potential is not straightforward, within the field range of $\pm 0.44,\si{V}/\si{\angstrom}$—which corresponds to similar values as in our study—the frequencies follow the same trend of increasing with negative polarization and decreasing with positive polarization of the metal surface. Overall, we notice that both metallic surfaces exhibit similar trends within the applied potential range, though Pd electrodes show more pronounced frequency variations for relatively smaller structural changes compared to Au.

\subsection{Water monolayer}

\begin{figure}[b!]
	\includegraphics[width=\columnwidth] {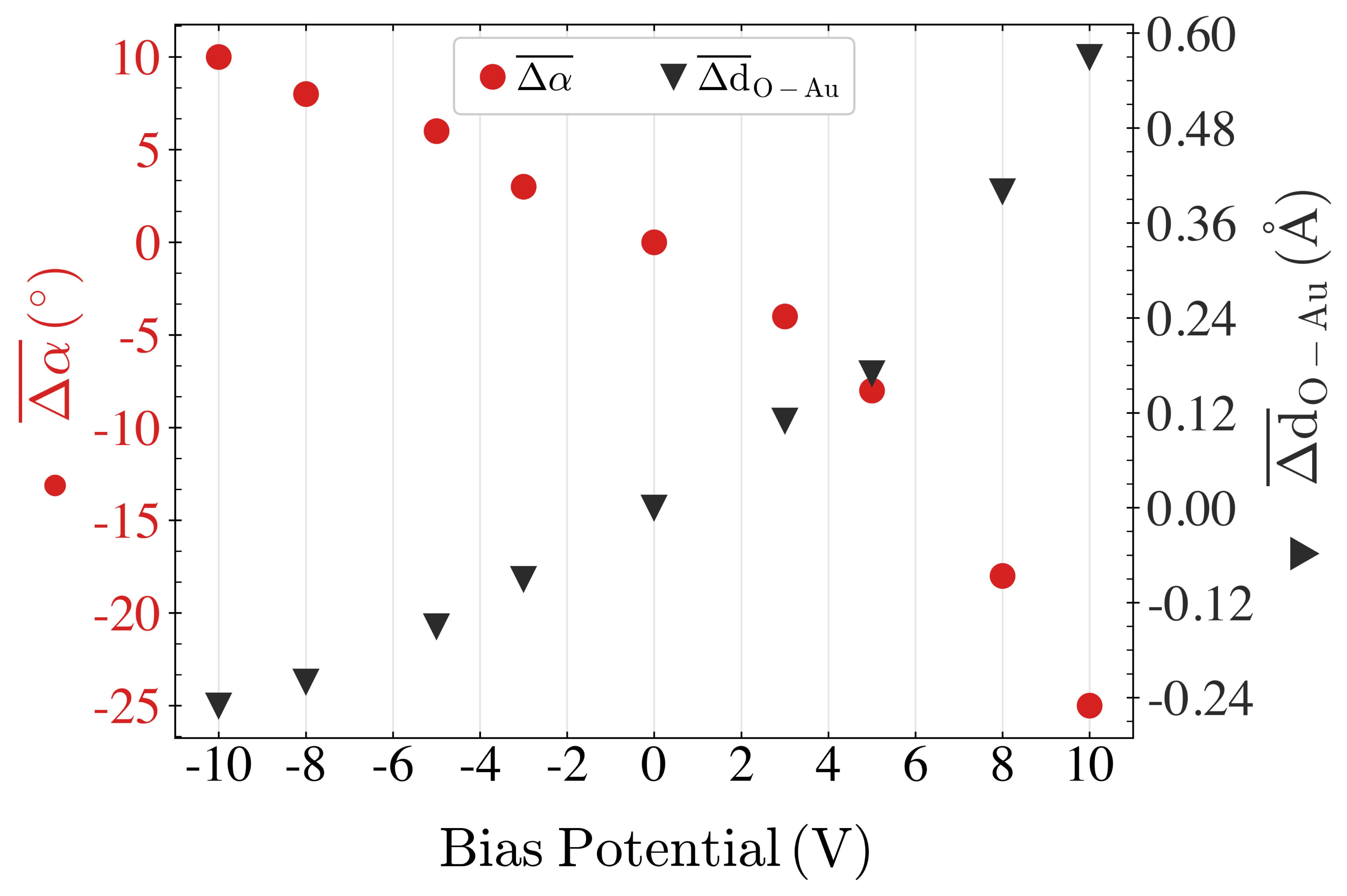}
	\caption{Averaged tilt angle ($\overline{\Delta \alpha}$) and vertical displacement ($\rm{\overline{\Delta d}_{O-Au}}$) variations of flat molecules in the minimum energy geometry of the water layer in each bias after the optimization.}
	\label{layer_geom}
\end{figure}

The monomer adsorbed on the metal surface can be viewed as a prototype system for understanding the water-metal interactions. However, SFG experiments actually probe water layers at the interface. Therefore, we also analyzed the vibrational frequencies of a water monolayer adsorbed on the Au(111) surface as a function of the applied bias.
As shown in Fig. \ref{fig:system}(d), the hexagons in this 2D layer are formed by alternating flat and H-down (dangling H atoms pointing towards the metal electrode) molecules. Under positive bias, the flat water molecules tend to move away from the metal surface, while under negative potentials, they move closer, as shown in Figure \ref{layer_geom}. The H-down molecules remain more tightly bound, with only minimal structural changes.  
The average maximum deviation for the tilt angle $\overline{\Delta\alpha}$ was just $3^{\circ}$, and for the vertical displacement $\overline{\Delta\rm{d}}_{\rm{O-Au}}$ was only $0.09\,\si{\angstrom}$ (see Figure S4 in the SI). 
This behavior of the flat water molecule is in line with the results observed for the monomer, though the variations in tilt angle and oxygen-metal distance are smaller in the monolayer. 
This can be attributed to the formation of hydrogen bonds between the water molecules.

\begin{figure}[t!]
	\includegraphics[width=.82\columnwidth]{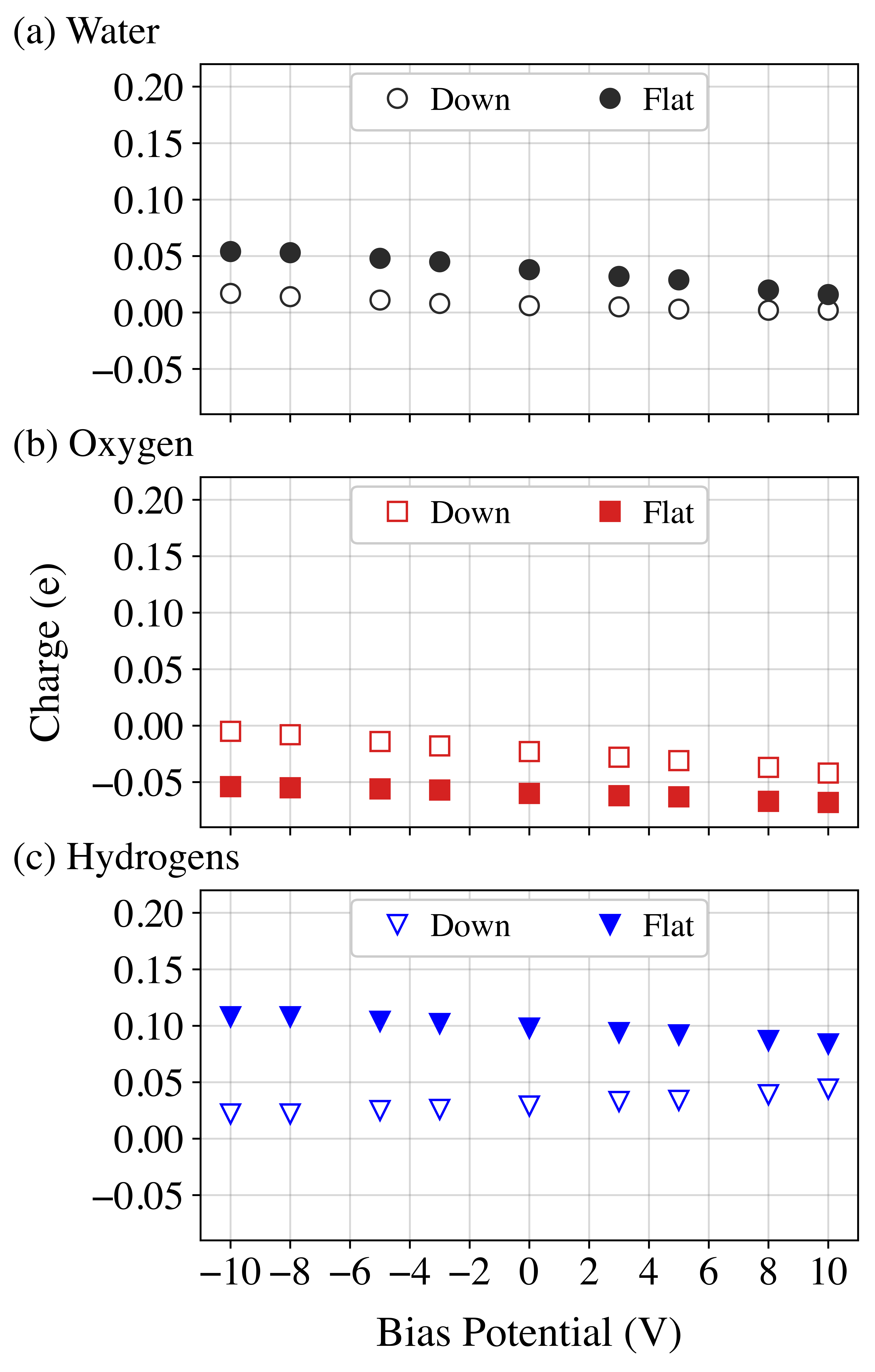}
	\caption{Averaged charge transfer of down (empty points) and flat (filled points) water molecules. (a) The black circles correspond to the total water charge; the (b) red squares and (c) blue triangles correspond to the O and H atom contributions.}
	\label{fig:mulliken_layer}
\end{figure}

\begin{figure}[b!]
	\includegraphics[width=0.9\columnwidth] {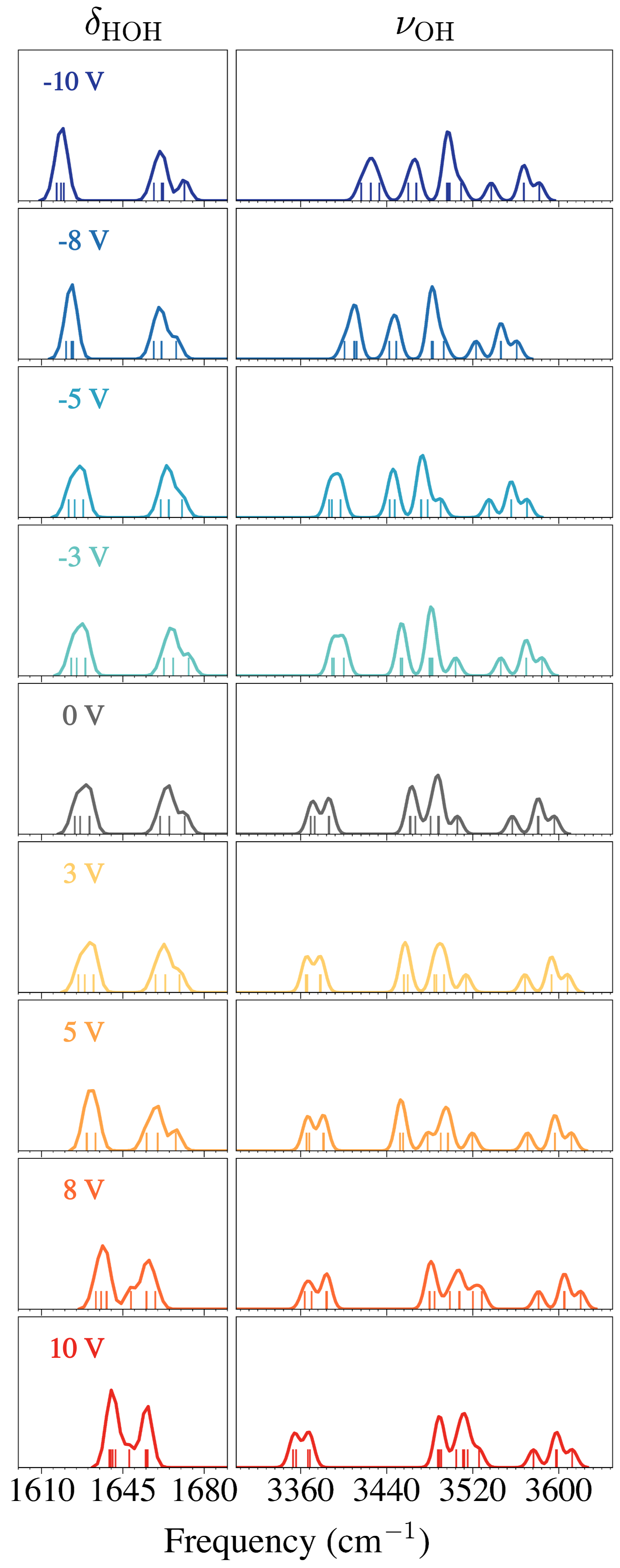}
	\caption{Normal mode frequencies of a water monolayer adsorbed on an Au(111) surfaces for different applied bias potentials. Left handside panels correspond to bending modes, and right handside panels to stretching modes.}
	\label{layer_freq}
\end{figure}

The asymmetric response of the system to the sign of the applied potential is still noted, and for negative potentials, there is a larger variation of the oxygen-oxygen distance ($\overline{\Delta\rm{d}}_{\rm{O-O}}$), whereas for positive bias the distances do not change as much with increasing potential (see SI, Figure S5). Furthermore, the effect of the external potential mostly affects the flat water molecules, as their interaction with the metal surface is stronger through the oxygen orbitals. This was also observed in \textit{ab-initio} molecular dynamics simulations on Pt(111) and Au(111) surfaces\cite{water_strucute_aimd,water_pt_aimd,potential_pzc_pt,coadsorption_aimd} using different techniques to try to mimic the external bias potential. In all cases, the interfacial water chemisorbs on the metal surface through the oxygen atom. In these studies, the authors also reported an excess of charge in the first water layer and a structural reordering of the subsequent water layers. 
However, the charge transfer observed in our calculations was small, indicating a weaker interaction between the water molecules and the metal electrode (see Figure \ref{fig:mulliken_layer}). Moreover, the $\rm{\sqrt{3}\times \sqrt{3}R30}$ water layer arrangement limits interaction between the metallic orbitals of non-atop site atoms and the water molecules. The presence of hydrogen bonds in the monolayer further restricts the influence of these orbitals, which could otherwise facilitate charge transfer (see SI, Figure S6).\cite{review-nature,electro_curcinotta}

Finally, the distribution of the bending and stretching modes is shown in Figure \ref{layer_freq}. To associate specific frequency values with their corresponding interactions, we analyzed the eigenvector directions for characteristic modes (see SI, Figure S8). The lower bending frequencies are associated to the flat-down molecules, while the higher-frequency vibrations are dominated by the flat-lying molecules. On the other hand, the lower stretching values correspond to the symmetric stretching modes of the molecules. For the negative potentials, the interaction between the flat-down molecules and the metal dominates these mode frequencies. As the bias increases, the frequencies correspond to the hydrogen bond interaction. Intermediate frequencies correspond to a combination of symmetric stretching modes of flat molecules and asymmetric stretching modes of flat-down molecules. The higher frequencies are associated with the asymmetric stretching vibrations of flat molecules.

Since there was no reorientation molecule, the stretching modes exhibit a monotonic behavior with bias, albeit with different trends. We observed that the lower frequencies tend to decrease with increasing bias, while the higher-frequency modes increase. These shifts, associated with the increase in the O-O distance of the flat-down molecule --acting as a hydrogen bond donor-- result in stronger hydrogen bonds and weaker water-metal interactions (see Figure S8), as we expected from the charge transfer analysis. These results are in line with the experimental observations using shell-isolated nanoparticle-enhanced Raman spectroscopy, where an increase in stretching frequencies was observed for interfacial water on negatively charged surfaces, while higher stretching values were reported for positively charged surfaces.\cite{raman2,experimental_pd} On the other hand, the bending frequencies show an opposite trend compared to the stretching ones, where we observe a separation of the peak frequencies for negative values and an assemble for positive potentials. In this sense, for positive potentials, we observed an increase in the bending frequencies associated with the flat-down molecules and a decrease in those of the flat molecules. This result suggests stronger hydrogen bonds and weaker water-metal interactions, as the molecules are further from the metal.\cite{bending}

\section{Conclusion}

In conclusion, we employed a combination of density functional theory (DFT) and non-equilibrium Green’s functions (NEGF) to investigate the structural and vibrational properties of a water molecule on Au(111) and Pd(111) surfaces, as well as a water monolayer on Au(111). Our approach allowed us to explicitly simulate a bias voltage drop within the grandcanonical ensemble, without the need of fictitious counter-electrodes.

For a single water molecule adsorbed on metallic surfaces, we found that under negative bias, the molecule moves closer to the surface, as the surface gets positively charged and interacts more strongly with the electronegative oxygen atom. Interestingly, we note that this does not result in significant charge transfer between surface and molecule, only charge rearrangements. Moreover, the interaction is not purely electrostatic, and a competition with Pauli repulsion arises. On the other hand, the molecule tends to rotate to the down configuration for positive bias voltages. The overall structural changes are more pronounced in Au compared to Pd, as the molecule is more strongly bound to the latter - and consequently, a higher charge transfer is seen. These structural changes were reflected on the vibrational frequencies with a non-monotonic behavior for the stretching modes. These changes reflect the molecule's rotation due to the charged surface, which, in turn, strengthens or weakens the O-H bonds. Thus, the molecule binds more strongly to the metal for positive potentials via the interaction between the H atoms and the negatively charged surface. On the other hand, for negative voltages, the H atoms move away from the positively charged surface and, as a result, interact less with the metal, leading to strengthened O-H bonds. 

In the case of the water monolayer on Au(111), the presence of hydrogen bonds limits large structural changes, with the external bias primarily affecting the oxygen-metal distance. The vibrational modes present can be separated into two families ascribed to flat and flat-down molecules, respectively. For the stretching modes the frequencies tend to get closer together as the bias is increased, whereas the opposite trend is noted for the bending modes. We also observed higher stretching frequencies at positive potentials, indicating that the O-H bonds in the flat-down molecules become weaker while the hydrogen bonds strengthen. Overall, our results suggest that the monolayer binds more strongly to the surface under negative bias and more weakly under positive bias, reflecting the dynamic interplay between water-metal interactions and hydrogen bonding at the interface.

Although the systems analysed in this work can be viewed as prototypes, this work provides new insights into the complex behavior of water adsorbed on electrified surfaces. Also, the configurations examined here can be used to train neural network-based force fields for modeling water/metal interfaces under bias, paving the way for more accurate and computationally efficient atomistic simulations.

\begin{acknowledgments}

The authors would like to acknowledge financial support from Funda\c{c}\~{a}o de Amparo \'a Pesquisa do Estado de S\~{a}o Paulo (FAPESP Grant \# 2017/10292-0, 2020/16593-4, 2017/02317-2, 2023/09820-2), Conselho Nacional de Pesquisa (CNPq) and CAPES. Calculations were carried out at CENAPAD-SP and at 
the Santos Dumont High performance facilities of Laboratorio Nacional de Computa\c{c}\~{a}o Cient\'{\i}fica (LNCC), Brazil.

\end{acknowledgments}

\section*{Data Availability Statement}

The data are available from the corresponding author upon resonable request. 

\nocite{*}

\bibliography{ref}
\bibliographystyle{apsrev4-2}
\end{document}



\title{Supplementary Information: Probing Water-Electrified Electrode interfaces: Insights from Au and Pd}

\author{Graciele M. Arvelos}
\affiliation{Instituto de F\'isica Te\'orica, Universidade Estadual Paulista (UNESP), S\~ao Paulo, SP 01140-070, Brazil}

\author{Marivi Fernández-Serra}
\affiliation{Physics and Astronomy Department, Stony Brook University. Stony Brook, New York 11794-3800, USA}
\affiliation{Institute for Advanced Computational Science, Stony Brook, New York 11794-3800, USA}

\author{Alexandre R. Rocha}
\affiliation{Max Planck Institute for the Structure and Dynamics of Matter, 22761 Hamburg, Germany}
\affiliation{Instituto de F\'isica Te\'orica, Universidade Estadual Paulista (UNESP), S\~ao Paulo, SP 01140-070, Brazil}

\author{Luana S. Pedroza}
\affiliation{Instituto de F\'isica, Universidade de S\~ao Paulo, SP 05508-090, Brazil}
\email{luana@if.usp.br}

\date{\today}

\maketitle

\onecolumngrid
\section{\label{sec:charge_transfer} Water-monomer/metal properties}

\subsection{Electronical Properties}
\begin{figure}[h!]
    \centering
	\includegraphics[width=0.95\columnwidth]{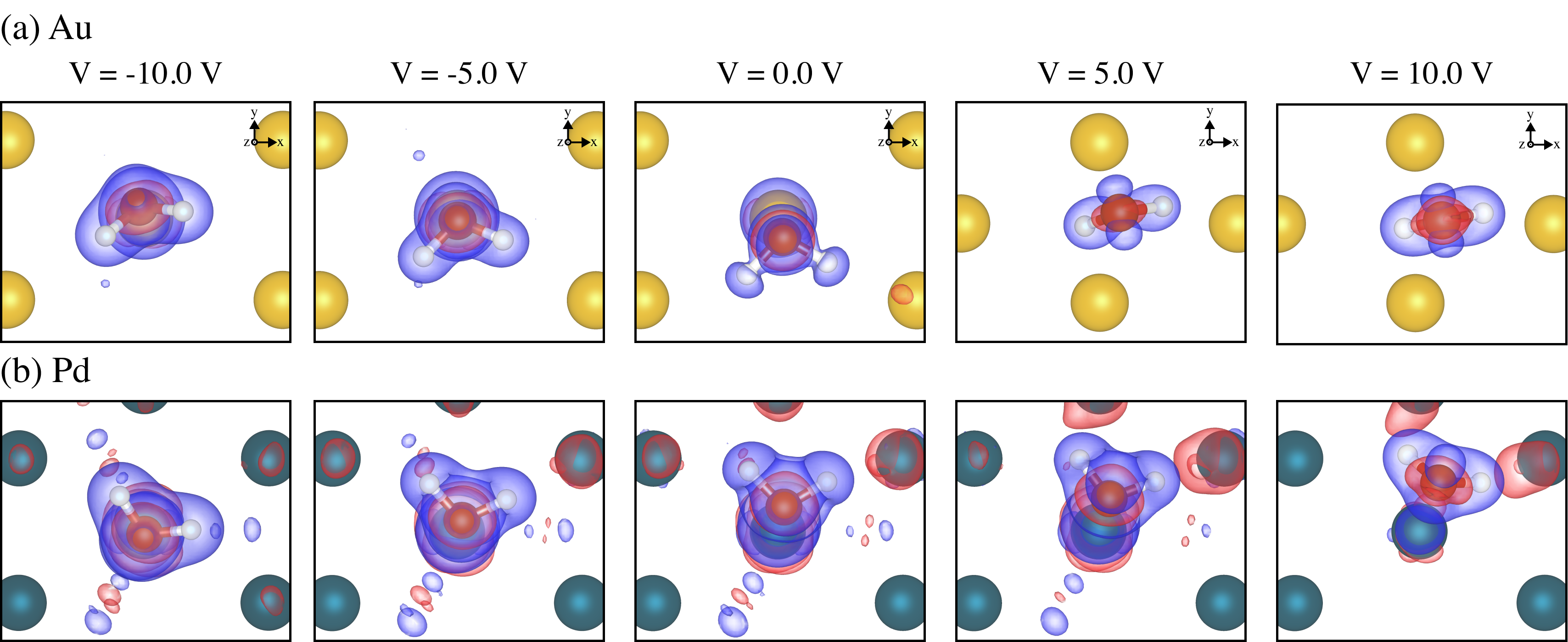}
    \caption{Plane view of the charge density difference between the system under an applied bias V and a combination of a parallel-plate capacitor with the same bias and an isolated water molecule subjected to an equivalent external electric field. The first row shows the monomer adsorbed on Au electrodes, while the second row corresponds to the Pd case. The isosurface value is $1.5\times10^{-3}\,\si{e}/\si{\angstrom}^3$ for all plots, with blue (red) indicating a decrease (increase) in electron desinty.}
	\label{fig:charge_density_up_SI}
\end{figure}
\begin{figure}[h!]
 \centering
	\includegraphics[width=0.95\columnwidth]{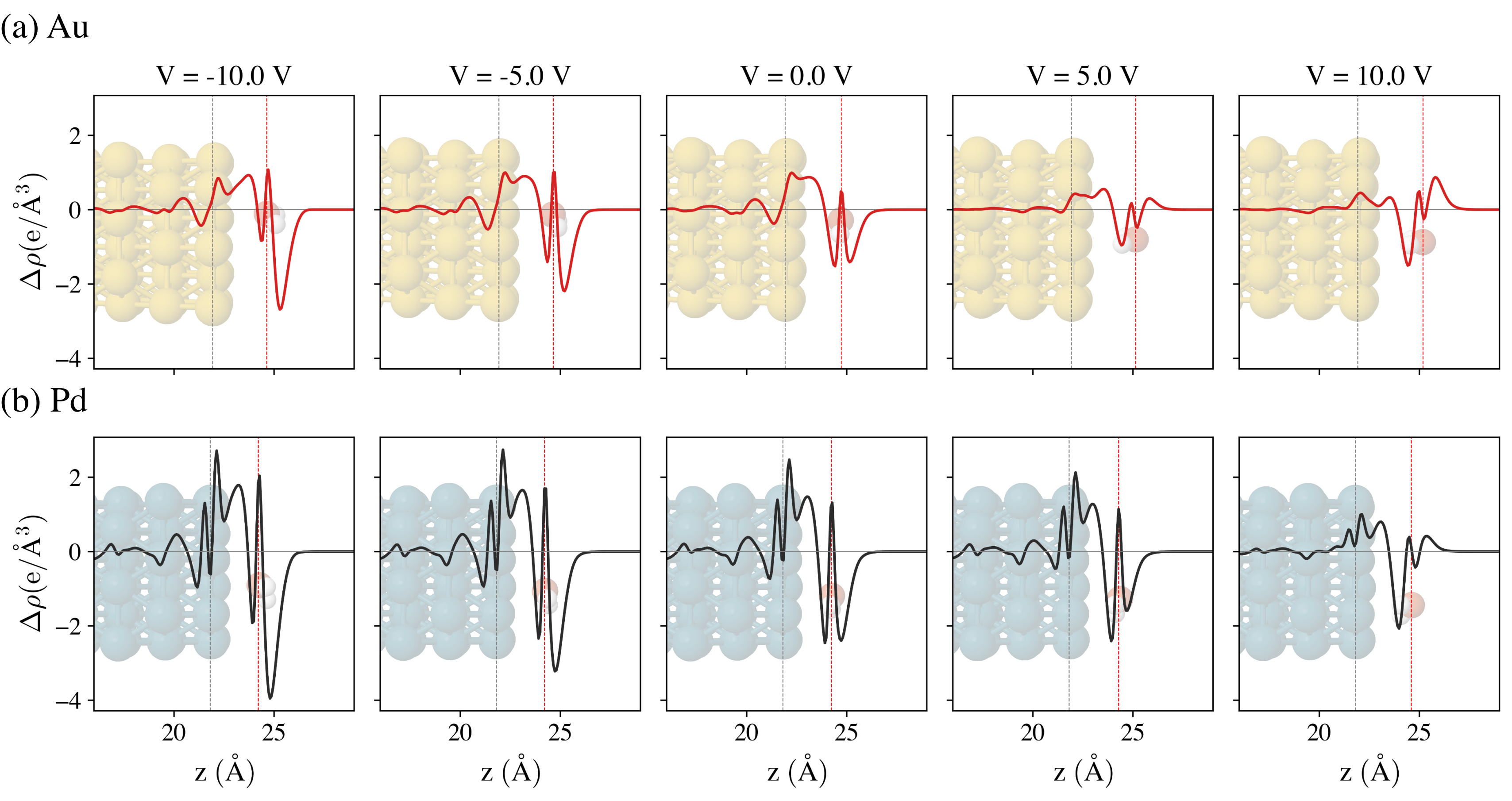}
    \caption{Difference in charge density averaged over the plane perpendicular to the z-axis. It was obtained by comparing the system under an applied bias V and a combination of a parallel-plate capacitor with the same bias and an isolated water molecule subjected to an equivalent external electric field. The dashed lines indicate the position of the last electrode layer (black) and the oxygen atom (red), respectively.}
	\label{fig:z_charge_SI}
\end{figure}

\subsection{Vibrational Properties}

\begin{figure}[H]
    \centering
	\includegraphics[width=\columnwidth]{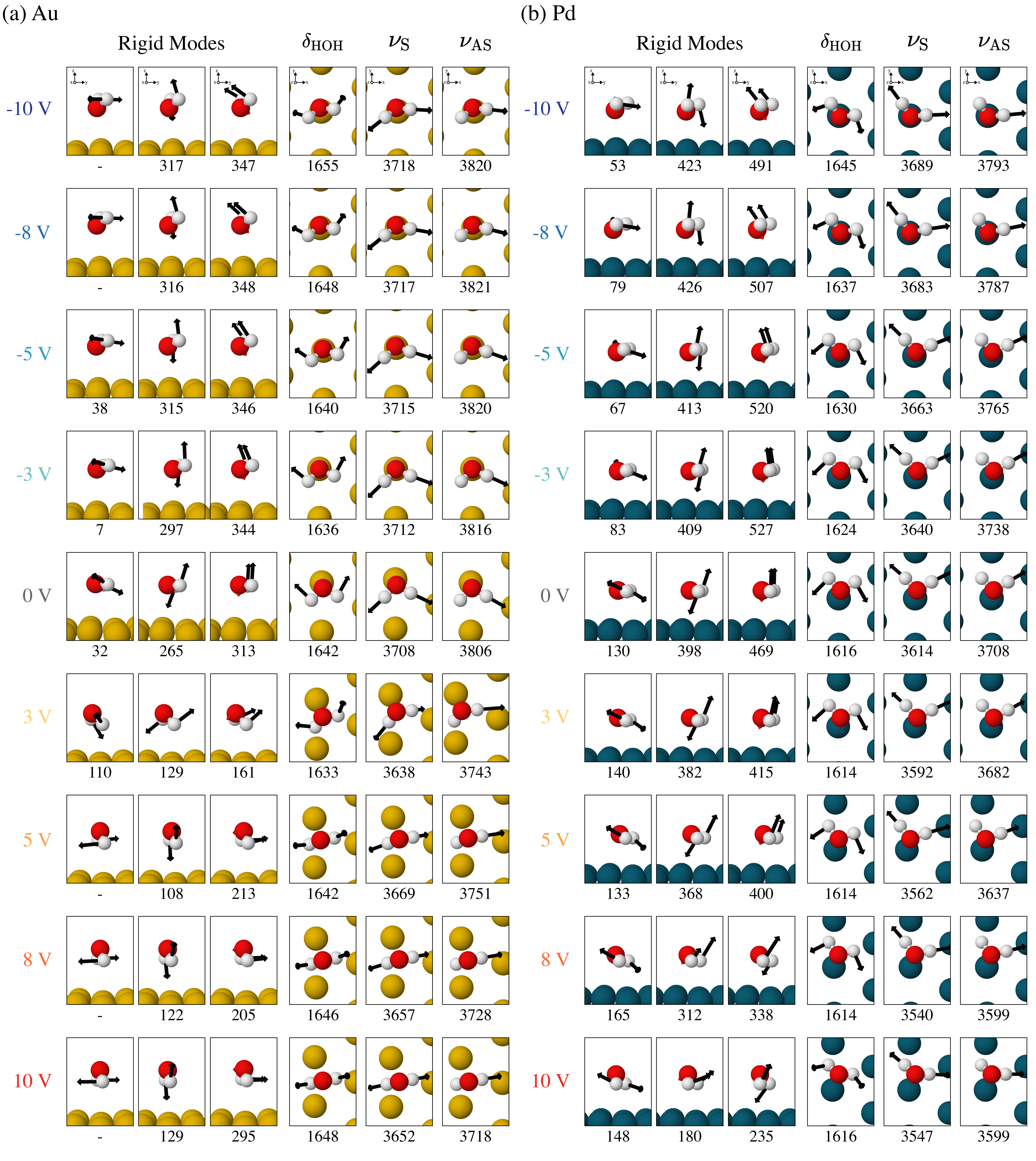}
    \caption{Visualization of the normal modes of vibration for the monomer adsorbed on (a) Au and (b) Pd surfaces under different applied bias potentials. Arrows indicate the direction of atomic vibrations.}
	\label{fig:modes_SI}
\end{figure}

To separate the structural and bias contributions to the vibrational properties, we calculated the normal mode frequencies at $V=0.0\si{V}$ using the optimized geometries obtained under different bias voltages, as shown in Table \ref{tab:frequencies_bias0}. When comparing the frequencies for the applied bias cases with those at zero bias, no significant differences were observed. This indicates that the external bias potential induces structural changes but does not directly affect the vibrational modes.

\begin{table}[H]
\centering
\caption{Frequencies of the bending normal mode ($\rm{\delta_{HOH}}$), and the symmetric ($\rm{\nu_S}$) and antisymmetric ($\rm{\nu_{AS}}$) stretching modes. The row for 0.0 V corresponds to the frequencies calculated using the optimized geometries obtained at the respective bias voltages.}
\label{tab:frequencies_bias0}
\begin{tabular*}{\columnwidth}{@{\extracolsep{\fill}}cc|*{3}{c}|*{3}{c}} 
\hline\hline
                     &       & \multicolumn{3}{c|}{Au}                                                                                   & \multicolumn{3}{c}{Pd}                                                                                     \\ 
\hline
\multicolumn{2}{c|}{Geometry}    & $\delta_{\rm{HOH}} ~(\si{\cm}^{-1})$ & $\nu_{\rm{S}} ~(\si{\cm}^{-1})$ & $\nu_{\rm{AS}} ~(\si{\cm}^{-1})$ & $\delta_{\rm{HOH}} ~(\si{\cm}^{-1})$ & $\nu_{\rm{S}} ~(\si{\cm}^{-1})$ & $\nu_{\rm{AS}} ~(\si{\cm}^{-1})$  \\ 
\hline
\cline{1-2}\cline{4-8}
\multirow{2}{*}{-10} & V = -10 & 1655                                                  & 3718                        & 3820                                             & 1645                                                  & 3689                        & 3793                          \\
                     & V = 0   & 1652                                                  & 3716                        & 3822                                             & 1643                                                  & 3687                        & 3793                          \\ 
\hline
\multirow{2}{*}{-8}  & V = -8  & 1648                                                  & 3717                        & 3821                                             & 1637                                                  & 3683                        & 3787                          \\
                     & V = 0   & 1647                                                  & 3715                        & 3822                                             & 1636                                                  & 3681                        & 3786                         \\ 
\hline
\multirow{2}{*}{-5}  & V = -5  & 1640                                                  & 3715                        & 3820                                             & 1630                                                  & 3663                        & 3765                          \\
                     & V = 0   & 1639                                                  & 3714                        & 3821                                             & 1629                                                  & 3661                        & 3763                          \\ 
\hline
\multirow{2}{*}{-3}  & V = -3  & 1636                                                  & 3712                        & 3816                                             & 1624                                                  & 3640                        & 3738                          \\
                     & V = 0   & 1636                                                  & 3712                        & 3816                                             & 1624                                                  & 3639                        & 3737                          \\ 
\hline
\multicolumn{2}{c|}{0.0}       & 1642                             & 3708                        & 3806                                             & 1616                                                 & 3614                        & 3708                          \\ 
\hline
\multirow{2}{*}{3}   & V = 3   & 1633                                                  & 3638                        & 3743                                             & 1614                                                  & 3592                        & 3682                          \\
                     & V = 0   & 1631                                                  & 3639                        & 3745                                             & 1614                                                  & 3594                        & 3683                          \\ 
\hline
\multirow{2}{*}{5}   & V = 5   & 1642                                                  & 3669                        & 3751                                             & 1614                                                 & 3562                        & 3637                          \\
                     & V = 0   & 1633                                                  & 3657                        & 3742                                             & 1615                                                  & 3564                        & 3642                          \\ 
\hline
\multirow{2}{*}{8}   & V = 8   & 1646                                                  & 3657                        & 3728                                             & 1614                                                  & 3540                        & 3599                          \\
                     & V = 0   & 1641                                                  & 3659                        & 3736                                             & 1613                                                  & 3544                        & 3608                          \\ 
\hline
\multirow{2}{*}{10}  & V = 10  & 1648                                                  & 3652                        & 3718                                             & 1616                                                  & 3547                        & 3599                         \\
                     & V = 0   & 1642                                                  & 3653                        & 3728                                             & 1612                                                  & 3551                        & 3612                          \\
\hline\hline
\end{tabular*}
\end{table}

\newpage   
\section{Water monolayer properties}
\subsection{Structural Properties}
\begin{figure}[h!]
\includegraphics[width=.67\columnwidth]{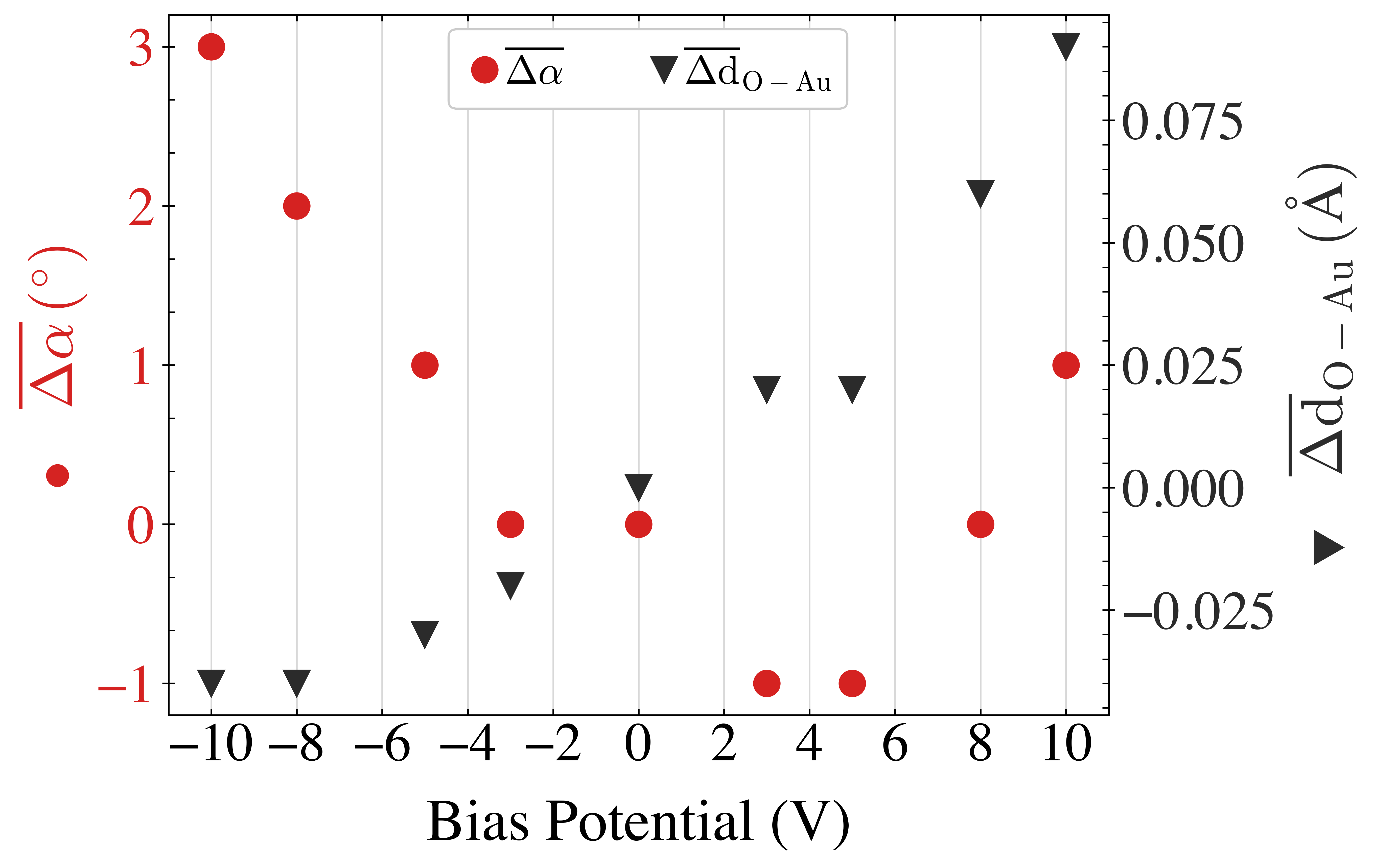}
    \caption{Averaged tilt angle ($\overline{\Delta \alpha}$) and vertical displacement ($\rm{\overline{\Delta d}_{O-Au}}$) variations of down molecules in the optimized minimum-energy geometry of the water layer at each applied bias.}
	\label{fig:layer_down_SI}
\end{figure}

\begin{figure}[!h]
    \includegraphics[width=.7\columnwidth]{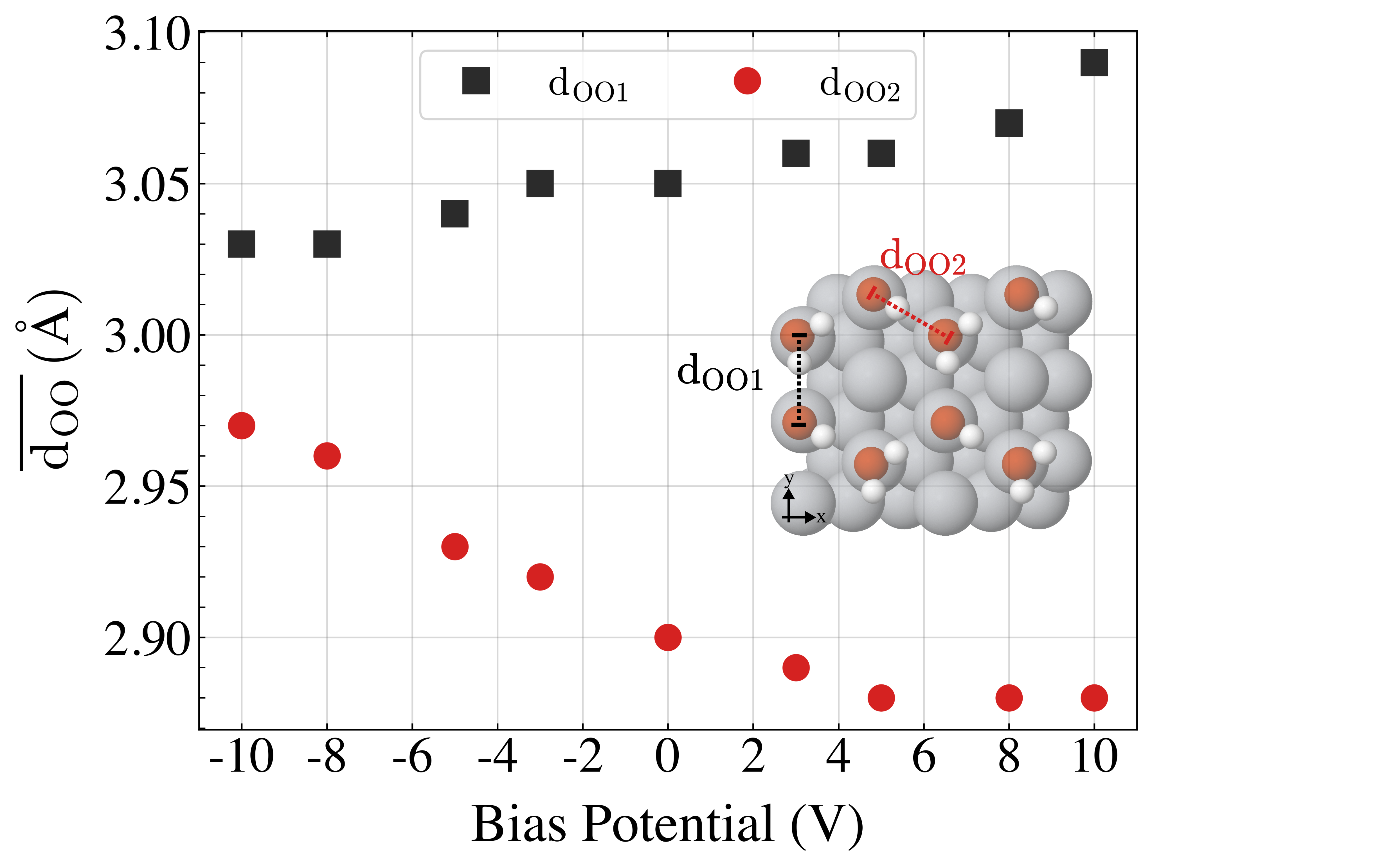}
    \caption{Averaged distances between oxygen atoms in the water layer under different applied biases. The distance between a flat water molecule acting as a hydrogen bond donor and a down molecule as the acceptor is denoted as $\rm{d_{OO1}}$, while the distance between a down molecule as the donor and a flat molecule as the acceptor is denoted as  $\rm{d_{OO2}}$.}
	\label{fig:layer_dOO_SI}
\end{figure}
\newpage
\subsection{Electronic Properties}

\begin{figure}[!ht]
	\includegraphics[width=\columnwidth]{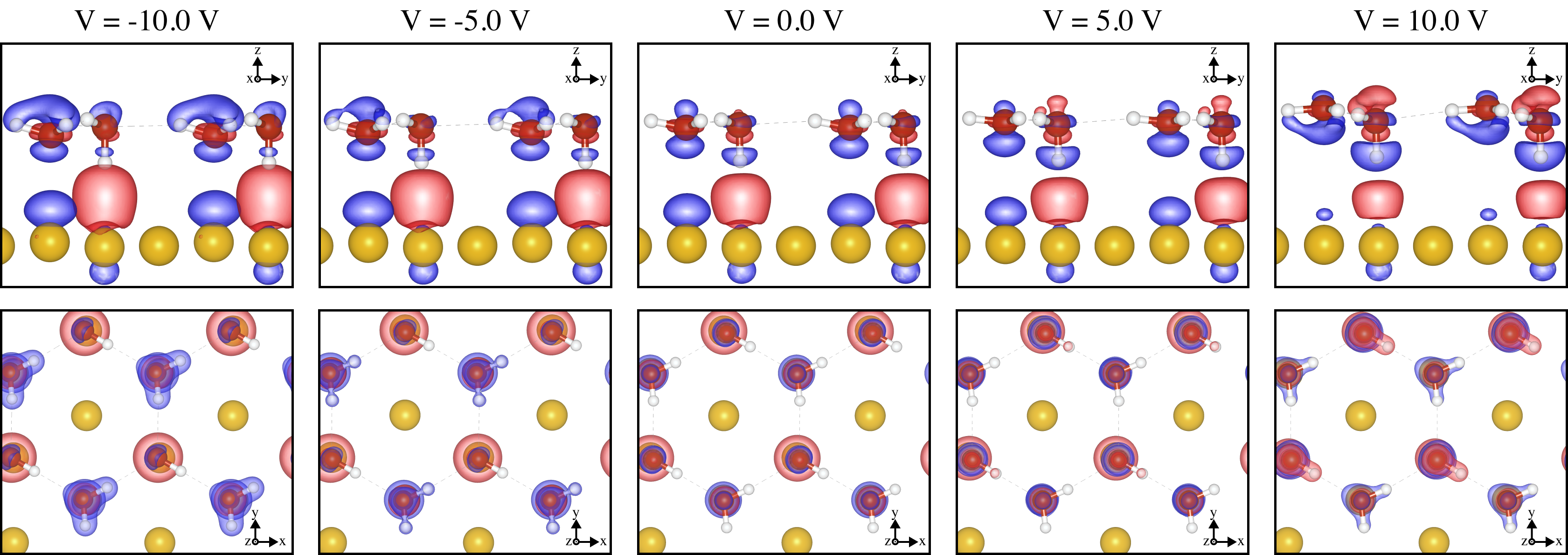}
    \caption{Plane and top views of the difference in charge density between the system under an applied bias V and a combination of a parallel-plate capacitor at the same bias with an isolated water layer subjected to an equivalent external electric field. The isosurface value is $1.5\times10^{-3}\,\si{e}/\si{\angstrom}^3$ for all plots, where blue (red) indicates a decrease (increase) in electron density.}
	\label{fig:charge_transfer_SI}
\end{figure}

\begin{figure}[!ht]
	\includegraphics[width=\columnwidth]{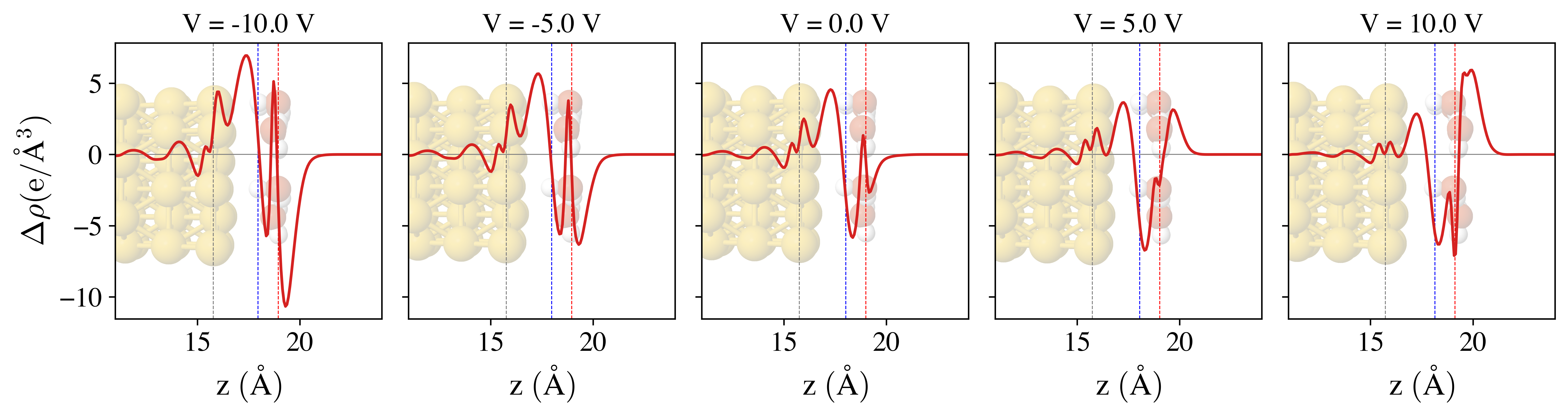}
    \caption{Summed charge density difference along the plane parallel to the z-axis.  It was obtained by comparing the system under an applied bias V with a combination of a parallel-plate capacitor at the same bias and an isolated water layer subjected to an equivalent external electric field. The dashed lines indicate the position of the last electrode layer, the hydrogens (blue), and the oxygens (red) of the down water molecule closest to the metal, respectively.}
	\label{fig:charge_z_SI}
\end{figure}
\newpage
\subsection{Vibrational Properties}

\begin{figure}[!ht]
	\includegraphics[width=\columnwidth]{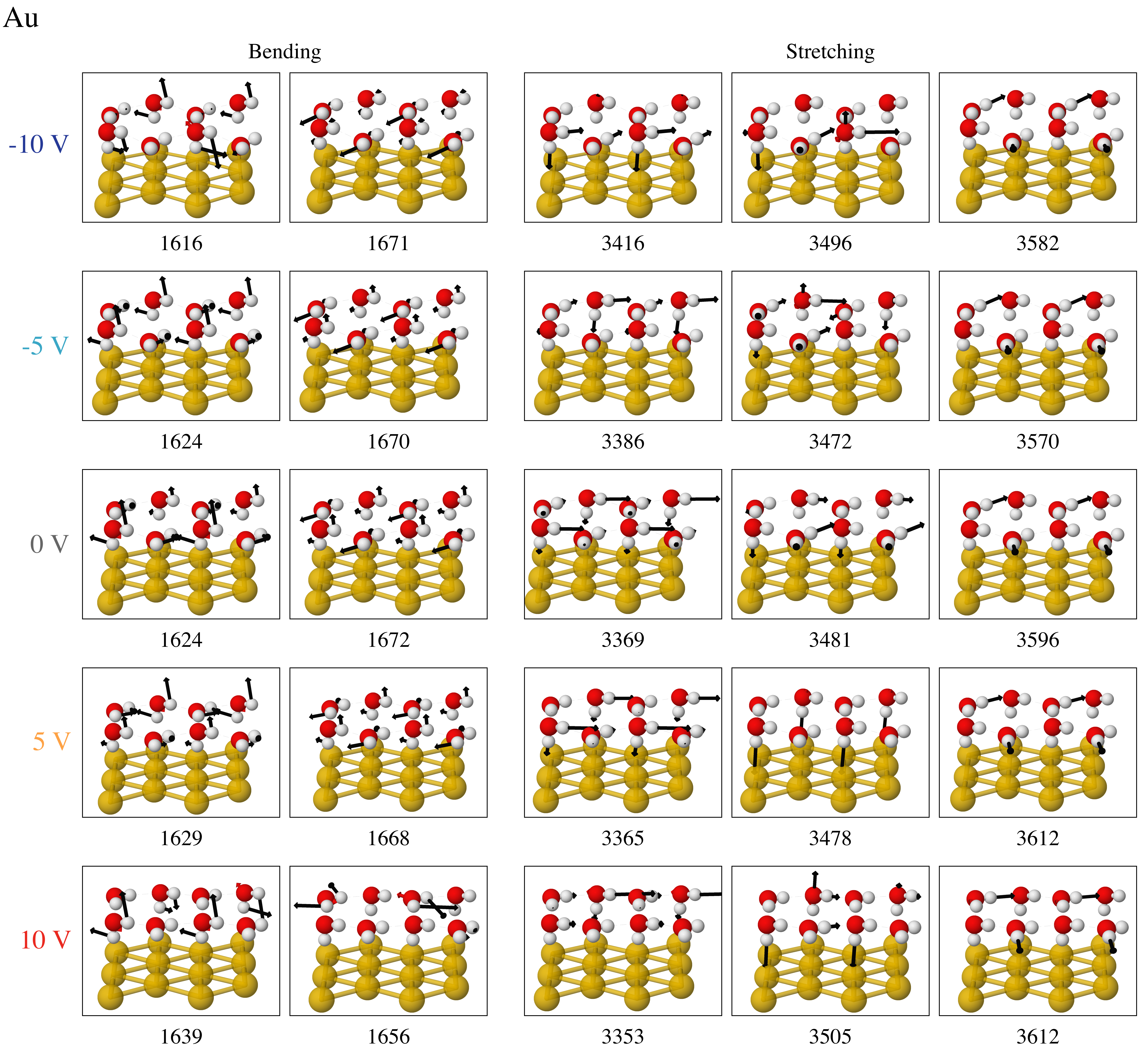}
    \caption{Visualization of the eigenvectors corresponding to the atomic vibrational frequencies of the water layer adsorbed on the Au electrode. The bending columns display the lowest and highest frequencies, while the stretching columns show the lowest, highest, and one representative intermediate frequency.}
	\label{fig:modes_layer_SI}
\end{figure}
